\documentclass[epj]{svjour}

\usepackage{epsfig,graphicx}

\begin{document}

\title{ Quark-diquark model for p($\bar{\textrm{p}}$)-p elastic scattering at high energies} 
\author{ 
V.M.~Grichine\inst{1}\thanks{Corresponding author, e-mail:  Vladimir.Grichine@cern.ch}
\and N.I. Starkov\inst{1}\and N.P. Zotov\inst{2}}
\institute{
Lebedev Physical Institute, Moscow, Russia \and
Skobeltsyn Institute of Nuclear Physics, Lomonosov Moscow State University, Moscow, Russia
}

\date{Received: date / Revised version: date}

\abstract{
A model for elastic scattering of protons at high energies based on the quark-diquark 
representation of the proton is discussed. The predictions of the model are compared 
with experimental data for the elastic  differential cross-sections from available databases. 
}

\maketitle

\section{Introduction}
\label{intro}

Recently the TOTEM Collaboration reported the first experimental data on the 
$pp$ elastic cross-section at a total energy in the center of mass system 
$\sqrt{s} = 7$ TeV (everywhere the Planck constant, $\hbar$, and the speed of light, 
$c$, are assumed to be unit, $\hbar=c=1$)~\cite{totem1}. Therefore there 
is opportunity to describe one in the more wide area of energy using the early data
(see, for example,~\cite{ZRT}). In general there are the number of models of the 
elastic $pp$ scattering description~\cite{ZRT,Dremin}.

In this note we consider the quark-diquark ($qQ$-) model in which 
baryons are considered as bound states of quark and diquark (a quasi-particle 
state of two quarks). This model appeared at the end sixties~\cite{LT} and 
was used for description of different problems: baryon spectroscopy~\cite{LT2}, 
multiparticale production~\cite{AK}, deep-inelastic processes
~\cite{BA} and go on. The  $qQ$-model was proposed by 
V.A. Tsarev~\cite{vat79} in 1979  to  describe the characteristics of  proton-proton 
elastic scattering and to explain the absence of second dip in the proton-proton 
differential cross-section, $d\sigma_{el}/dt$, at the four-momentum transfer $-t\sim 4-5$~GeV$^2$, 
which should exist if a proton would compounded of tree quarks. 

A. Bialas and A. Bzdak reinvented in 2007~\cite{bb2007} the approach of~\cite{vat79} in more 
simplified form without the real part of the scattering amplitude. They showed that the 
quark-diquark model is capable to predict the correct position of the proton-proton 
differential elastic cross-section minimum, however, the absence of the scattering amplitude 
real part results in dramatic overestimation of the value of first dip at $-t\sim 1.3$~GeV$^2$. 
The model~\cite{bb2007} was applied in~\cite{nc2012} to describe at the same level of accuracy 
the TOTEM data~\cite{totem1}  for $d\sigma_{el}/dt$ at $\sqrt{s}=$7~TeV.
 
Below we recall the main features of the $qQ$-model~\cite{vat79} suitable for numerical 
calculations and provide comparison with the experimental data on the $pp$ elastic differential  
cross-section in the region of high energies, $\sqrt{s}\geq$546~GeV.

\section{Quark-diquark model for p($\bar{\textrm{p}}$)-p elastic scattering}

The proton-proton differential elastic cross-section can be expressed in terms of the scattering 
amplitude $F(s,t)$:
\begin{equation}
\label{dsdt}
\frac{d\sigma_{el}}{dt}=\frac{\pi}{p^2}|F(s,t)|^2, 
\end{equation}
where $p$ is the proton momentum in the center of mass system. 

The model~\cite{vat79} limits the consideration of the scattering amplitude by contributions from one- 
and two-pomeron exchanges between quark-quark (1-1), diquark-diquark (2-2) and 
quark-diquark (1-2). In this approximation  $F(s,t)$ reads:
\begin{equation}
\label{f}
F(s,t)=F_1(s,t)-F_2(s,t)-F_3(s,t),
\end{equation}
where $F_1(s,t)$ is the scattering amplitude with one-pomeron exchange, while $F_2(s,t)$ 
and $F_3(s,t)$ correspond to two-pomeron exchanges between the proton constituents, 
quark and diquark. The amplitude $F_1(s,t)$ reads:
\[
F_1(s,t)=\frac{ip\sigma_{tot}(s)}{4\pi}\left[B_1\exp(A_{11}\,t)+B_2\exp(A_{22}\,t)+\right.
\]
\begin{equation}
\label{f1}
\left.+2\sqrt{B_1B_2}\exp(A_{12}\,t)\right],
\end{equation}
where $\sigma_{tot}(s)$ is the total  
proton-proton cross-section. The coefficients $B_1$, and $B_2$ parametrize the quark-quark, 
$\sigma_{11}$, and the diquark-diquark, $\sigma_{22}$, cross-sections, respectively:
\[
\sigma_{11}=B_1\sigma_{tot}(s), \quad\sigma_{22}=B_2\sigma_{tot}(s).
\]
The model assumes the quark-diquark cross-section, $\sigma_{12}=\sqrt{\sigma_{11}\sigma_{22}}$. 

The coefficients $A_{jk}$, ($j,k=1,2$) are derived taking into account the Gauss distribution of quark and 
diquark in proton together with the standard Pomeron parametrization. They read ($s_0$=1~GeV$^2$):
\[
A_{jk}=\frac{r^2_j+r^2_k}{16}+\tilde{\alpha}\left[\ln\frac{s}{s_o}-\frac{i\pi}{2}\right]+
\]
\begin{equation}
\label{Ajk}
+\lambda\left[\left(\frac{m-m_j}{m}\right)^2+\left(\frac{m-m_k}{m}\right)^2\right].
\end{equation}
Here  $r_1$, $m_1$ are the quark radius and mass, and  $r_2$, $m_2$ are 
the diquark radius and mass, respectively; $\tilde{\alpha}$=0.15~GeV$^{-2}$ is the 
Pomeron trajectory slope, and $\lambda=r^2/4$, where $r$ is the proton radius. In the
model it is assumed that the $m_1=m/3$ and the $m_2=2m/3$, where $m$ is the proton mass. 
The $r_1$ and $r_2$ were found by the fitting of experimental data: $r_1=0.173\,r$, $r_2=0.316\,r$. 

The amplitudes $F_2(s,t)$ and $F_3(s,t)$ are:
\[
F_2(s,t)=\frac{ip}{4\pi}\frac{B_1B_2\,\sigma^2_{tot}(s)}{8\pi(A_{12}+4\lambda/9)}\cdot
\]
\[
\cdot\left[\exp\left(\frac{A_{11}A_{22}-(4\lambda/9)^2}{2(A_{12}+4\lambda/9)}\,t\right)\right.+
\]
\begin{equation}
\label{f2}
\left.
+\exp\left(\frac{A_{12}-4\lambda/9}{2}\,t\right)\right],
\end{equation}
and:
\[
F_3(s,t)=\frac{ip}{4\pi}\frac{\sqrt{B_1B_2}\,\sigma^2_{tot}(s)}{4\pi}
\left[\frac{B_1}{A_{11}+A_{12}-4\lambda/9}\cdot\right.
\]
\[
\cdot\exp\left(\frac{A_{11}A_{12}-(2\lambda/9)^2}
{A_{11}+A_{12}-4\lambda/9}\,t\right)+
\]
\begin{equation}
\label{f3}
\left.+\frac{B_2}{A_{12}+A_{22}+2\lambda/9}\exp\left(\frac{A_{12}A_{22}-(\lambda/9)^2}
{A_{12}+A_{22}+2\lambda/9}\,t\right)\right],
\end{equation}
respectively. The quark-quark cross-section, $\sigma_{11}$ and the proton radius $r$ are the free 
parameters defining (together with $\sigma_{tot}(s)$) the $s$-dependence of the $d\sigma_{el}/dt$. 
The  diquark-diquark cross-section and the parameter  
$B_2$ are derived from the optical theorem, which results in the 
following equation:
\[
\sigma_{tot}b_1B_1B_2+\sigma_{tot}\sqrt{B_1B_2}(b_2B_1+b_3B_2)=
\]
\begin{equation}
\label{B2}
=B_1+B_2+2\sqrt{B_1B_2}-1,
\end{equation}
where
\[
b_1=\frac{1}{4\pi}\textrm{Re}\left[\frac{1}{A_{12}+4\lambda/9}\right],\quad
\]
\[
b_2=\frac{1}{4\pi}\textrm{Re}\left[\frac{1}{A_{11}+A_{22}-4\lambda/9}\right],
\]
\[
b_3=\frac{1}{4\pi}\textrm{Re}\left[\frac{1}{A_{12}+A_{22}+2\lambda/9}\right].
\]
Equation~(\ref{B2}) is the third-order equation relative to $\sqrt{B_2}$. For $0<B_1<1$, 
it has rational root $0<\sqrt{B_2}<1$ which is used in the model.

%%%%%%%%%%%%%%%%%%%%% Elastic ds/dt %%%%%%%%%%%%%%%%%%%%%%%%%%%%%%

\begin{figure}
\includegraphics[height=2.8in,width=3.5in]{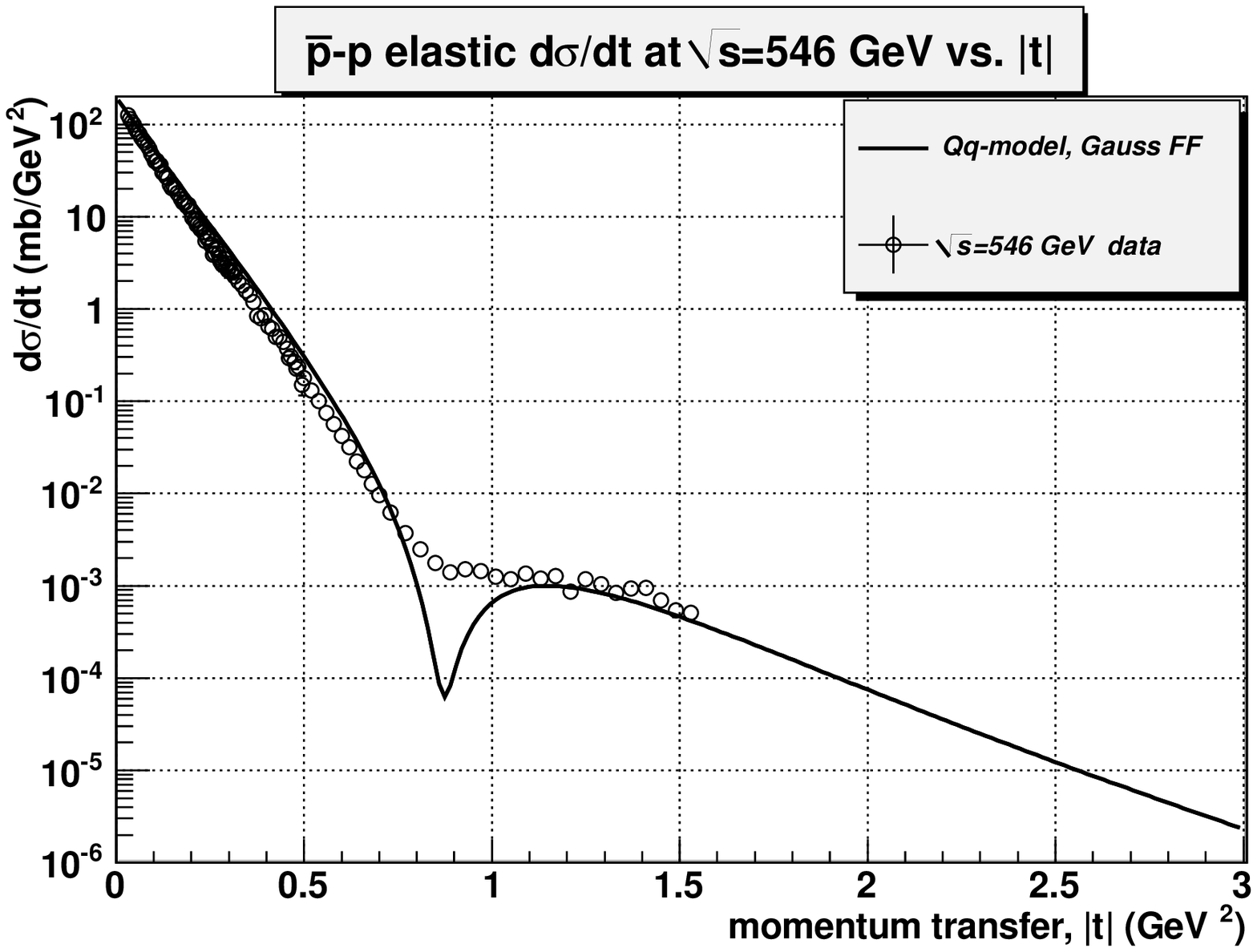}
\caption{The antiproton-proton differential elastic cross-section 
versus $|t|$ at $\sqrt{s}=$546~GeV. The curve is the prediction of our model. 
The open circles are the experimental data~\cite{pbp546}. }
\label{546gev}
\end{figure}

\begin{figure}
\includegraphics[height=2.8in,width=3.5in]{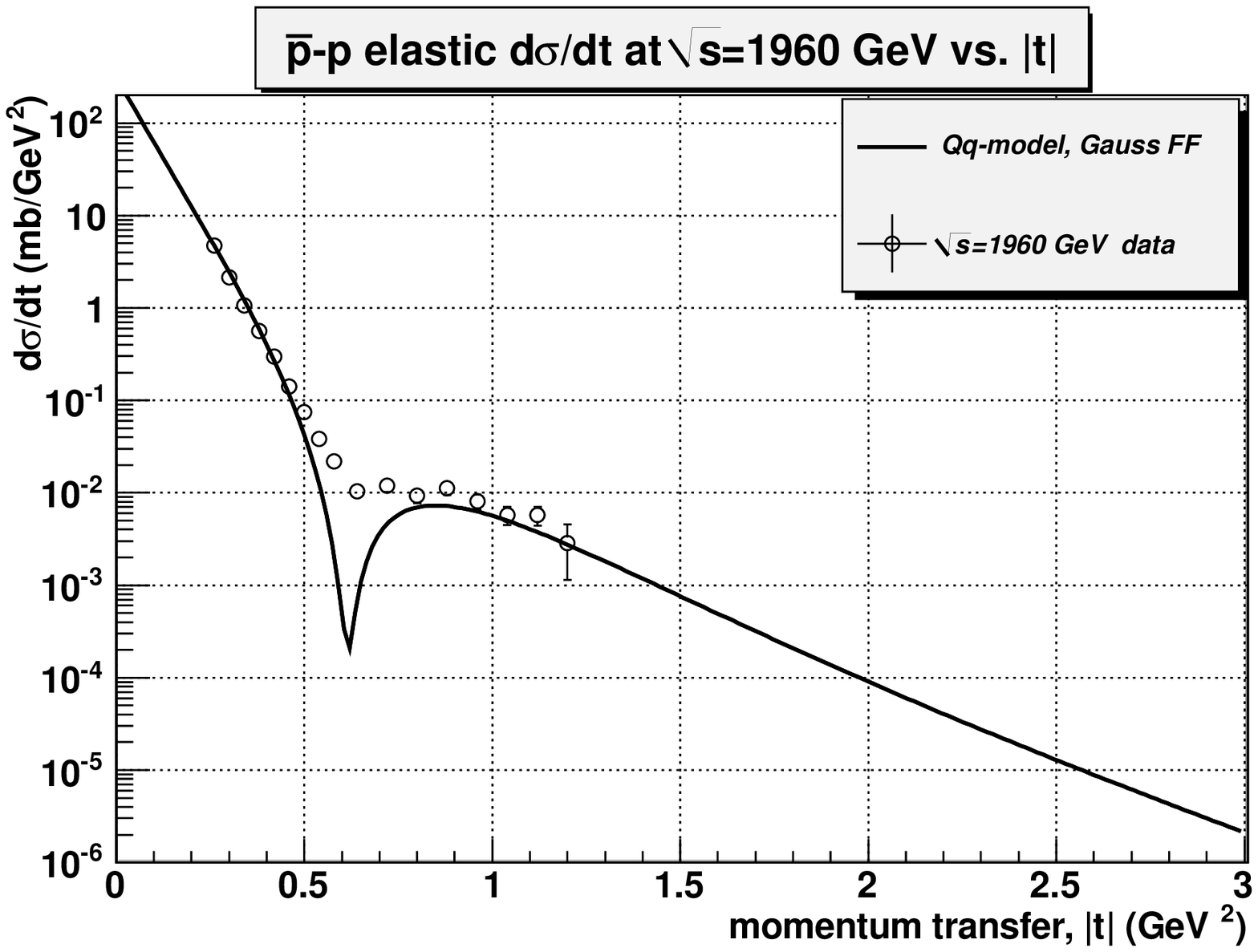}
\caption{The antiproton-proton differential elastic cross-section 
versus $|t|$ at $\sqrt{s}=$1960~GeV. The curve is the prediction of our model. 
The open circles are the experimental data~\cite{pbp1960}. }
\label{1960gev}
\end{figure}

\begin{figure}
\includegraphics[height=2.8in,width=3.5in]{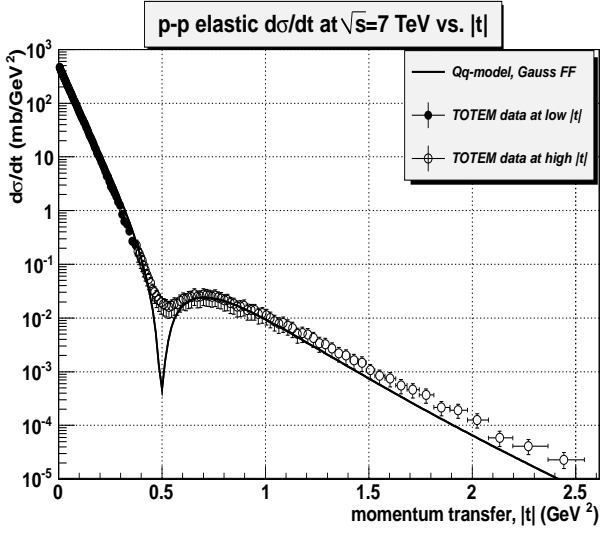}
\caption{The proton-proton differential elastic cross-section 
versus $|t|$ at $\sqrt{s}=$7~TeV. The curve is the prediction of our model. 
The open and closed circles are the LHC TOTEM experimental data 
from~\cite{totem1}. }
\label{7tev}
\end{figure}

%%%%%%%%%%%%%%%%%%%%%%%%%%%%%%%%%%%%%%%%%%%%%%%%%%%%%%%%%%%%%%%%%%%%%%%%%%%%%%%%%%%%%

\section{Comparison with experimental data}

Fig.~\ref{546gev},\ref{1960gev},\ref{7tev} show the antiproton-proton differential elastic 
cross-section versus $|t|$ at $\sqrt{s}=$546~GeV, $\sqrt{s}=$1960~GeV, and the proton-proton 
differential elastic cross section at $\sqrt{s}=$7~TeV, respectively.  The curves are the 
predictions of our model. We see that the proposed $qQ$-model describe reasonably the 
differential elastic cross sections of the elastic antiproton-proton and proton-proton scattering
in  wide region of energy.

%%%%%%%%%%%%%%%%%%%%%%%%%%%%%%%%%%%%%%%%%%%%%%%%%%%%%%

\begin{figure}
\includegraphics[height=2.8in,width=3.5in]{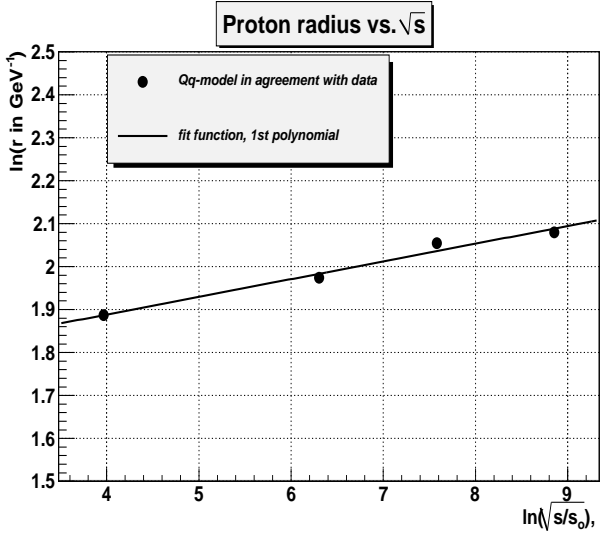}
\caption{ The dependence $\ln{(r)}$, $r$ in GeV$^{-1}$, versus the $\ln{(\sqrt{s/s_0})}$.}
\label{Rpvs}
\end{figure}

\begin{figure}
\includegraphics[height=2.8in,width=3.5in]{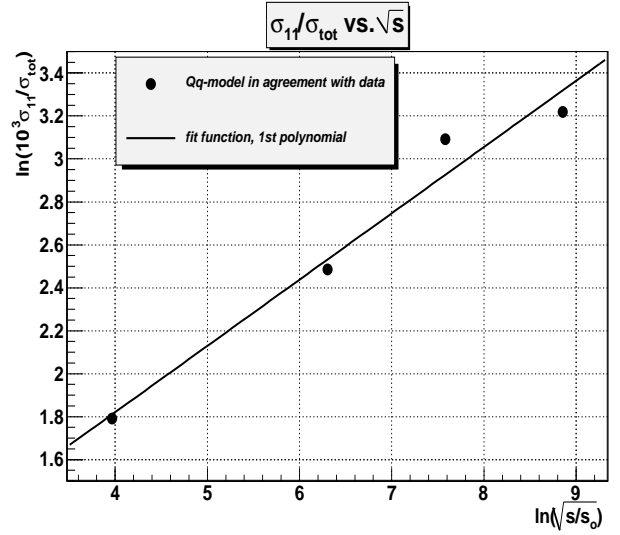}
\caption{ The dependence $\ln{(10^3\sigma_{11}/\sigma_{tot})}$ versus $\ln{(\sqrt{s/s_0})}$.}
\label{B1vs}
\end{figure}

%\section{Discussion and Summary}
The results shown in fig.\ref{546gev}-\ref{7tev} correspond to the $s-$dependence of the model 
free parameters which are shown in fig.\ref{Rpvs}-\ref{B1vs}. The model results in increase 
of the proton radius and the quark-quark cross-section with growth of $s$.

\section{Discussion and Summary}

We have considered the $qQ$-model of the $pp$-elastic scattering at high energies. We have obtained reasonable
description of  the differential cross section of elastic $pp$ scattering in wide region of energies.
The position of the $d\sigma_{el}/dt$-minimum is in for the  satisfactory agreement with experimental data, while the
value of  dip is overestimated, though less than in the model~\cite{bb2007}. The reason is that the 
model~\cite{vat79} has nonzero real part of the scattering amplitude coming from the Pomeron 
parametrization. However, the value of the real part is not enough for correct description of the dip value. 

The model tuning with more sophisticated form-factors, investigation of additional sources for 
the scattering amplitude real part, and more broad comparison of the model with experimental data are current plans.

\section*{Acknowledgment}
 
We are thankful to S. Giani for fruitful discussions of the paper contents.
N.Z. is  grateful to the DESY Directorate for the support within the Moscow -- DESY 
project om Monte-Carlo implementation for HERA -- LHC, he was also supported by FASI of 
Russian Federation (grant NS-3920.2012.2), the RFBR State contract 02.740.11.0244 and 
the Ministry of Education and Science of Russian Federation under agreement No. 8412. 
V.G and N.S. were partly supported by the CERN -- RAS Program of Fundamental Research at LHC.

%%%%%%%%%%%%%%%%%%%%%%%%%%%%%%%%%

\end{document}